# Carrier-Envelope-Offset Frequency Stabilization of a High Peak and Average Power Thin-Disk Oscillator

**Yasmin Kopp,\* Gregor Hehl, Johann Gabriel Meyer, Simon Goncharov, and Oleg Pronin**

*Helmut Schmidt University, Holstenhofweg 85, 22043, Hamburg, Germany*
*\*yasmin.kopp@hsu-hh.de*

**Abstract:** In this work, we demonstrate carrier-envelope-offset (CEO) frequency stabilization of a Kerr-lens mode-locked thin-disk oscillator delivering 180 W average power, 80 MW output peak power, and >500 MW intra-cavity peak power – the highest peak power achieved in any stabilized thin-disk oscillator system. The CEO frequency detection is performed with an f-2f interferometer based on supercontinuum generation in a YAG crystal. Intra-cavity loss modulation using an acousto-optic modulator, that simultaneously provides the Kerr lens, yields 50 mrad of residual phase noise with a 250 kHz control bandwidth. After pulse compression to 0.9 GW peak power in a dual-stage multipass cell, the system directly enables high harmonic generation in noble gases. These results represent a significant step in realizing a compact and robust high-repetition-rate driver system suitable for vacuum ultraviolet and extreme ultraviolet frequency comb generation.

## 1. Introduction

Accessing the vacuum ultraviolet (VUV) and extreme ultraviolet (XUV) spectral regions with a coherent source remains a significant challenge in modern physics. While existing light sources in this range - such as synchrotron radiation [1] - offer the necessary photon energies, they are rather complex systems. Similarly, laser-driven systems using enhancement cavities, capable of generating coherent light in this spectral region, are relatively complex in design [2]. One particular promising application which is currently being explored in this spectral region is the direct excitation of the 1S-2S transition of $He^+$ ions, which requires a coherent source in the deep UV around a wavelength of 60.8 nm [3]. Likewise, recent advances toward a new generation of ultra-precise nuclear clocks have demonstrated the thorium nuclear transition, lying around 148 nm [4]. One of the most straightforward ways to reach these spectral regions is frequency up-conversion via high-harmonic generation (HHG) in a gas jet driven by an existing near-infrared laser. However, because the efficiency of such a nonlinear process is only on the order of about $10^{-6}$ [5], a technology to increase the VUV/XUV output power is required. To avoid complex amplification stages, a laser driver with exceptionally high average power is essential. The phase noise is amplified during HHG [6], which can be highly detrimental when performing high-precision spectroscopy. Due to relatively low repetition rates, the use of a solid-state gain medium with the combination of Kerr-lens mode-locking, and high intra-cavity powers, thin-disk oscillator systems are expected to have significantly reduced intensity noise [7]. Due to amplitude-to-phase coupling, the carrier-envelope-offset (CEO) frequency and phase noise are expected to be correspondingly low. When combined with external spectral broadening and pulse compression, Kerr-lens mode-locked high-power oscillators can generate ultrashort pulses approaching the few-cycle regime [8,9]. CEO frequency stabilization of these systems represents a key milestone toward realizing compact, robust driving sources for HHG.

To detect the CEO frequency, a self-referencing scheme, such as an f-2f interferometer, is required. An f-2f scheme usually requires supercontinuum generation covering at least one octave, which is routinely performed in photonic crystal fibers or waveguides [10,11]. However, supercontinuum generation in fibers tends to degrade spatial and temporal coherence for pulses with durations longer than 100 fs [10]. Alternatively, white light can be generated in bulk media [12] via femtosecond filamentation [13,14], where the spatial and temporal coherence is preserved [15]. As an advantage, the alignment process is simpler and requires fewer optical components compared to photonic crystal fiber approaches, as demonstrated for CEO frequency stabilization in [16], using a thin-disk oscillator with an average output power of 28 W. Several CEO frequency stabilization methods have been demonstrated in high power laser oscillators. These techniques typically employ modulation of the pump power [17,18], the intra-cavity losses [19], or the implementation of feed-forward control schemes [20]. Among these, pump power modulation is particularly attractive for high power systems, including thin-disk oscillators, due to its intrinsic scalability. However, the achievable control bandwidth (several tens of kHz for most solid-state lasers) in such schemes is a fundamental limitation. By incorporating phase lead filters in the phase-locked loop (PLL), the bandwidth of pump power modulation can be extended to higher frequencies [21], enabling successful CEO frequency stabilization, as reported for a thin-disk oscillator with an average output power of 45 W in [22]. Furthermore, the CEO frequency stabilization of low average output power (<10 W) thin-disk oscillators was reported in various works [23,24], achieving relatively low residual noise performances of 120 mrad [23].

Intra-cavity loss modulation increases the control bandwidth compared to pump power modulation and has been experimentally shown in various systems: fiber and bulk oscillators using electro-optic modulators [19,25], thin-disk oscillators utilizing acousto optic modulators (AOM) [8,26], and bulk solid-state lasers incorporating cw-pumped semiconductor saturable absorber mirrors (SESAMs) as opto-optical modulators [27]. To date, CEO frequency stabilization of Kerr-lens mode-locked thin-disk oscillators via intra-cavity loss modulation has only been realized in systems with a maximum output peak power of 67 MW and, more critically, with intra-cavity peak powers around 200 MW [26]. In that previous implementation using an AOM for CEO frequency control, a residual in-loop phase noise of 90 mrad was reported. However, extending this stabilization approach to higher power oscillators presents significant challenges. Achieving higher intra-cavity powers necessitates careful selection of the Kerr medium to avoid optical damage and nonlinear distortion. Additionally, higher intra-cavity power levels demand either a stronger cooling system - which induces vibrations of the thin-disk or must manage increased thermal instabilities inside the oscillator.

Feed-forward stabilization of the CEO frequency has proven effective in low-power, Ti:sapphire oscillators [20]. However, scaling this technique to high power laser systems present several intrinsic challenges. This method relies on the first diffraction order from an AOM, which demands high diffraction efficiency. For high-power beams, the beam diameter must be significantly enlarged to mitigate nonlinear effects such as self-focusing and thermal lensing within the AOM. This enlargement severely limits both, diffraction efficiency and modulation bandwidth [26]. Moreover, effective feed-forward control necessitates high passive CEO frequency stability to avoid beam-pointing instabilities in the first diffraction order [28]. These constraints substantially complicate the system design, often requiring the addition of feedback loops [29], and severely diminish the practicability for high power lasers.

Here, we used a method for CEO frequency stabilization where the AOM simultaneously acted as the Kerr medium and as the intra-cavity loss modulator [26]. A large modulation bandwidth was achieved by positioning the AOM within the intra-cavity focusing telescope (see Fig. 1). The self-focusing induced by the Kerr-effect in the AOM is used to initiate and stabilize the mode-locking process itself [30]. Using this approach, we achieve CEO frequency stabilization with a residual in-loop phase noise of 50 mrad (integrated from 1 MHz - 1 Hz) at an unprecedented average output power of 180 W, corresponding to 80 MW output peak power and an intra-cavity peak power exceeding 500 MW. Furthermore, our laser oscillator system delivers up to 0.9 GW peak power after pulse compression [9]. This achieves the highest average and peak power reported to date for a CEO frequency-stabilized laser oscillator system. Compared to [26], this represents more than an order-of-magnitude increase in output peak power for stabilized and compressed oscillator systems, while the residual in-loop phase noise was reduced from 90 mrad to 50 mrad. These results were achieved using an all-solid-state octave-spanning spectrum, generated through compression down to 55 fs in a multipass cell (MPC) followed by efficient supercontinuum generation in YAG via filamentation.

## 2. Setup and experimental results

The experimental setup is shown in Fig. 1. The Kerr-lens mode-locked Yb:YAG thin-disk oscillator used here is a slightly modified version of the oscillator described in previous work [30]. The AOM, made of crystalline quartz with a thickness of ~5 mm was used as a Kerr medium, placed under Brewster's angle in the vicinity of the focus of the telescope. With a telescope length of 2.5 m and a focus of around 400 µm, it operated with a focal spot almost a factor of two smaller compared to previously reported CEO frequency stabilized thin-disk oscillator [26]. The 5-mm AOM used here was operating in the Raman-Nath regime and with an estimated diffraction efficiency of max. 4 % at 4 W RF power. The beam was positioned roughly 1 mm below the transducer. The thin-disk oscillator was set to emit an average output power of 180 W, delivering 140 fs pulses with pulse energy around 14 µJ at a repetition rate of 14 MHz and a central wavelength of 1032 nm. For practical reasons, the output power compared to previous work [30] was reduced. It should be noted that when using the AOM as a Kerr medium at full pump power, the oscillator delivered the same peak- and average-output power levels (110 MW, 202 W) as presented in [30].

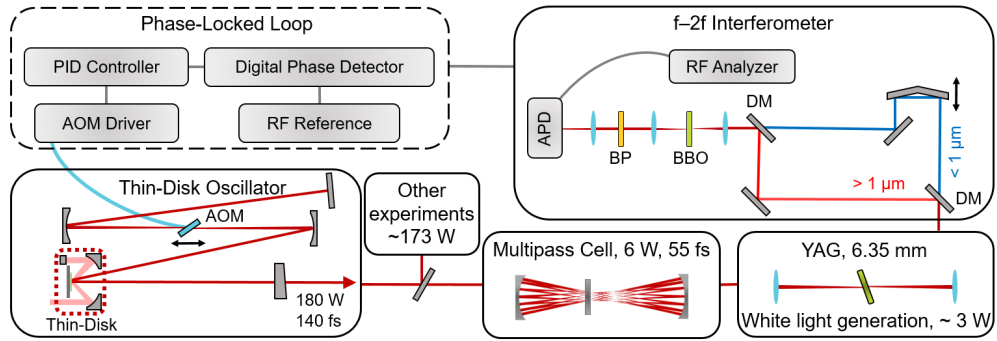

**Fig. 1.** *Schematic setup for the CEO frequency detection and stabilization: Thin-disk oscillator with AOM as Kerr medium; MPC compression stage; white light generation in YAG, f-2f interferometer layout with DM, dichroic mirror; BBO, beta barium borate for second harmonic generation; BP, bandpass filter; APD, avalanche photodetector and electronics for the PLL.*

The AOM was installed to control the CEO frequency via intra-cavity loss modulation [8]. During the experiment, a maximum RF power of 1 W was applied to the AOM. The AOM's performance, when implemented as a loss modulator, was characterized by measuring the transfer function of the oscillator output. This was achieved by manually applying a modulated signal to the AOM and recording the resulting intensity-modulated oscillator output (response) with an oscilloscope (Fig. 2). A phase lag of 90° and a 3 dB decrease in the modulation intensity were reached at around 250 kHz (marked in red in Fig. 2), defining the available CEO frequency control bandwidth. The amplitude transfer function exhibited a noticeable reduction in modulation depth for frequencies below the inverse of the upper-state lifetime of the gain medium (approximately 1 ms for Yb:YAG), as observed and discussed previously for intra-cavity loss modulation by the authors of reference [19]. However, the low amplitude response for frequencies <1 kHz can be compensated within the servo control loop.

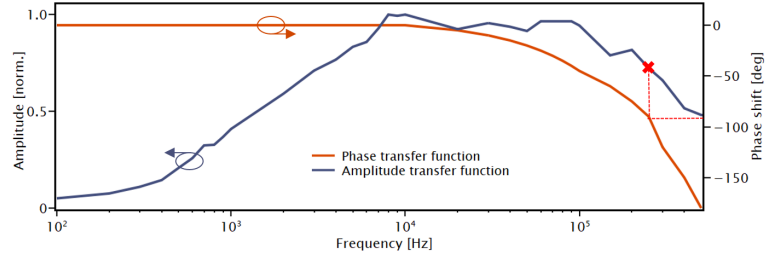

***Fig. 2.*** *Phase and amplitude transfer functions of the oscillator output power in response to the AOM modulation measured with a digital oscilloscope. The dashed red line indicates the frequency at which a phase shift of 90° is accumulated, and the modulation bandwidth of the AOM is achieved.*

While the laser oscillator was running at a full average power of 180 W, only an attenuated beam from an output coupler with an average power of 7 W was used for the supercontinuum generation and the subsequent CEO frequency stabilization. Thereby, it was possible to use in parallel most of the output power for other pulse compression and HHG experiments [9,31]. In order to generate a filament for supercontinuum generation in YAG, we first employed a low-power Herriott-type MPC to compress the laser pulses temporally, and thereby increase the peak power. Approximately 7 W and 0.5 µJ of the oscillator output underwent nonlinear spectral broadening in the bulk MPC. The MPC consisted of two dispersive concave mirrors (ROC = 50 mm) separated by 95 mm with a 6.35 mm-thick AR-coated fused silica window placed in the middle (see Fig. 1). This configuration led to a B-Integral of ~0.4 rad per pass. After 34 passes through the fused silica plate, the chirped pulses were compressed down to 55 fs (the estimated Fourier limit is 50 fs) with a set of dispersive mirrors providing a total GDD of -2000 fs². The pulse compression was verified with a commercial autocorrelator assuming Gauss-shape of the compressed pulses (Fig. 3(a, inset)).

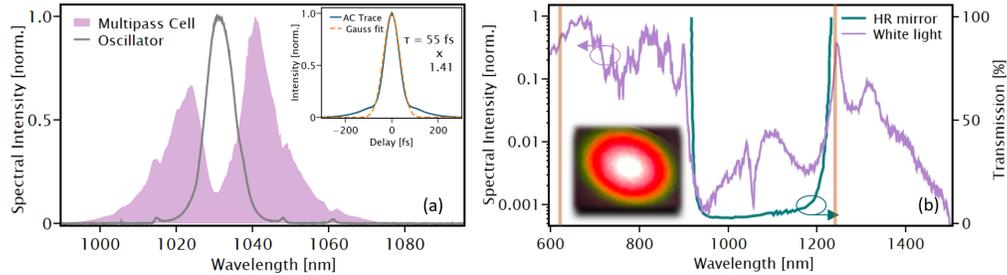

***Fig. 3.*** *(a) Spectra at the output of the oscillator (grey) and at the output of the Herriott-type MPC (violet) measured with an optical spectrum analyzer. Inset: autocorrelation trace measured at the output of the chirped-mirror compressor after the MPC. (b) Output spectrum after supercontinuum generation using filamentation in YAG crystal (violet) measured with an optical spectrum analyzer and transmission curve of the HR mirror (cyan). The orange lines indicate the spectral components used for beating in the f-2f interferometer. The inset shows a photo of the supercontinuum beam made by a smartphone camera.*

As a next step, the compressed 55-fs-long pulses were guided into an undoped 6.35 mm-thick YAG plate for supercontinuum generation through filamentation. The input power to the YAG plate was limited to ~3 W, and the resulting output spectrum covered a spectral range of 600-1450 nm (see Fig. 3(b)). The orange lines indicate the spectral components used for the f-2f interferometer. The dip in spectral intensity around 1000 nm arises from the use of a dielectric mirror with high reflectivity in this region, positioned after the YAG crystal (cyan curve in Fig. 3(b)). This mirror was intentionally implemented to suppress the high power spectral components near 1000 nm, which are not required for CEO frequency detection. The supercontinuum light was then coupled into the f-2f interferometer.

The schematic of the f-2f interferometer is illustrated in Fig. 1. The CEO frequency detection was performed using a self-built f-2f interferometer based on a Mach-Zehnder geometry. This layout was chosen to enable the implementation of spectral filtering within the interferometer arms, which was crucial to mitigate saturation in the APD. Spectral separation was achieved using dichroic mirrors, isolating the long wavelengths of the supercontinuum, which were subsequently long-pass filtered at 1150 nm. The filtered red part was recombined with the blue spectral components, which passed through a variable delay line to ensure temporal overlap between the fundamental and second harmonic components on the APD. The second harmonic generation was realized in a BBO crystal. A 620 nm bandpass filter (10 nm FWHM) was used before the detection to further minimize saturation on the APD by suppressing the contribution from non-overlapping spectral components.

The beating signal between the two interferometer arms was aligned and subsequently optimized to reach 40 dB above the noise floor. The electronic components of the phase-locked loop (PLL) for the CEO frequency stabilization are illustrated in Fig. 1. Before locking, the pump current of the laser oscillator was adjusted to shift the free-running beat signal to 10.7 MHz. A combination of bandpass filtering and a 50 dB gain stage was implemented before comparing it with a reference signal (10.7 MHz, Tabor Electronics LS3082B), using a $32\pi$ digital phase detector (Menlo Systems DXD200). The resulting error signal was sent to a proportional-integral-derivative (PID) controller (Vescent Photonics D2-125), while the monitoring output was used for displaying the locking performance on an oscilloscope. The servo output from the PID controller was used to modulate the driver signal of the intra-cavity AOM, thereby controlling the CEO frequency. Optimizing the control loop parameters by setting the corner frequencies of the first integrator to 50 kHz, the second integrator to 2 kHz, the derivative part to 100 kHz, and appropriately tuning the proportional gain, a tight phase lock of the CEO frequency was achieved (see Fig. 4(a)).

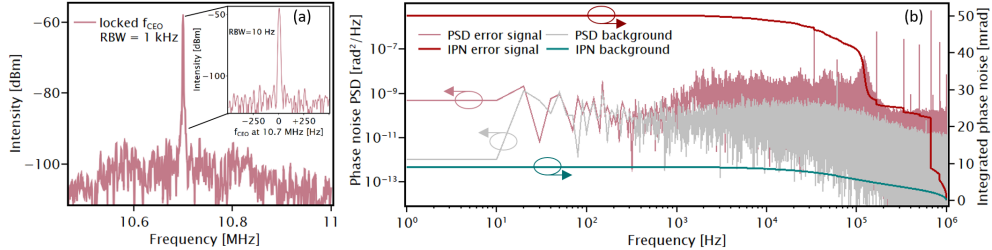

*Fig. 4.* (a) Locked CEO frequency to a reference, measured with a resolution bandwidth (RBW) of 1 kHz. Inset shows the linewidth, instrument-limited to a 10 Hz RBW. (b) Residual in-loop phase noise. The light red curve shows the power spectral density (PSD) of the CEO beat assessed from the recorded error signal. The integrated phase noise (IPN) from 1 MHz to 1 Hz is displayed in dark red. The noise floor PSD of the electronic loop, including the digital phase detector, is shown in grey and the corresponding IPN in cyan.

The free-running CEO frequency was detected using an RF spectrum analyzer with an RBW of 1 kHz. The unlocked CEO frequency experienced excursions up to 80 kHz (RMS), as discussed previously in [32]. When the CEO frequency was locked, the main peak was narrowed to sub-10 Hz (FWHM), limited by the RBW of 10 Hz (see inset in Fig. 4(a)). However, a pedestal, spanning around 300 kHz, persisted around the main peak with a difference of 10 dB to the noise floor. Furthermore, when stabilizing the CEO frequency, frequency peaks around the locked CEO frequency at 10.7 MHz appeared. Further investigation of the stabilized CEO frequency was done by analyzing the power spectral density (PSD).

The time trace of the error signal between the CEO frequency and the RF reference from the digital phase detector was recorded using a digital oscilloscope and analyzed to calculate the PSD and integrated phase noise (IPN). The residual in-loop phase noise was determined to be 50 mrad (1 MHz - 1 Hz). Fig. 4(b) displays the PSD (light red) and the corresponding IPN (dark red) of the error signal. The phase detector background noise is shown in grey (PSD) and cyan (IPN). The residual IPN of the background was estimated to be ~9 mrad. The PSD exhibits a spike at 50 Hz and its harmonics associated with electronics contributions. The narrow peaks at high frequencies over 150 kHz, as well as the broad peak around 120 kHz, were pronounced in the PSD, which contributed most to the IPN. Detailed noise analysis (PLL, phase detector, oscilloscope) is underway.

The YAG crystal used for supercontinuum generation combined with a pre-compression in MPC, employed in this work, required no realignment during operation. Despite the supercontinuum being generated through filamentation, no difference in amplitude fluctuations was observed when comparing the pulse train directly to the oscillator output using a photodiode. A stable CEO frequency operation was maintained for over 30 minutes. This demonstration confirms the approach's scalability to high output average- and peak powers as discussed in previous works [26,33]. However, compared to [26], our configuration employs the AOM simultaneously as both the Kerr medium and the intra-cavity loss modulation element, without requiring an additional Kerr medium. We thereby achieved a residual in-loop phase noise reduced by approximately a factor of two compared to reference [26], while simultaneously demonstrating output average and peak powers of 180 W and 80 MW, respectively.

## 3. Conclusion

In this work, we have demonstrated CEO frequency stabilization of a Kerr-lens mode-locked thin-disk oscillator operating at unprecedented output peak- (80 MW) and average (180 W) powers, along with an intra-cavity peak power exceeding 500 MW, with a residual phase noise of 50 mrad. Moreover, the laser system is capable of delivering 0.9 GW peak power in the compressed few-cycle regime [9]. An AOM was used simultaneously as a Kerr medium and for intra-cavity loss modulation. Thereby, a large control bandwidth of 250 kHz at a phase lag of 90° in the transfer function was achieved. To perform these experiments, we realized a f-2f interferometer based on the white light generation through filamentation in a YAG medium. We have previously demonstrated [9,31] that the compressed output of this oscillator system can directly drive high-harmonic generation (HHG) in noble gases, achieving 7th harmonic generation at full oscillator power without additional amplification stages. This demonstrates the immediate practical applicability of the system as a compact, robust, all-solid-state driver for next-generation frequency comb spectroscopy.

The final technical step toward a high power frequency comb source is repetition-rate stabilization [34], which is actively underway alongside optimization toward a simplified 2f-3f self-referencing scheme using high-power compressed output from a dual stage MPC. These advancements will provide a full phase stability, enhanced long-term robustness, and improved noise performance, enabling unprecedented precision in VUV/XUV frequency comb spectroscopy applications.

**Acknowledgments.** We thank Nazar Kovalenko for the technical support with the modification of the AOM and Moinuddin Kadiwala for his support with the multipass cell configuration.

**Disclosures.** The authors declare no conflict of interest.

**Data availability.** Data underlying the results presented in this paper are available on request.

## References

1. N. de Oliveira, M. Roudjane, D. Joyeux, D. Phalippou, J.-C. Rodier, and L. Nahon, "High-resolution broad-bandwidth Fourier-transform absorption spectroscopy in the VUV range down to 40 nm," Nat. Photon. **5**, 149–153 (2011).
2. P. Rupper and F. Merkt, "Intense narrow-bandwidth extreme ultraviolet laser system tunable up to 20 eV," Review of Scientific Instruments **75**, 613–622 (2004).
3. J. Moreno, F. Schmid, J. Weitenberg, S. G. Karshenboim, T. W. Hänsch, T. Udem, and A. Ozawa, "Toward XUV frequency comb spectroscopy of the 1 S–2 S transition in He$^+$," Eur. Phys. J. D **77**, 67 (2023).
4. C. Zhang, T. Ooi, J. S. Higgins, J. F. Doyle, L. von der Wense, K. Beeks, A. Leitner, G. A. Kazakov, P. Li, P. G. Thirolf, T. Schumm, and J. Ye, "Frequency ratio of the 229mTh nuclear isomeric transition and the 87Sr atomic clock," Nature **633**, 63–70 (2024).
5. S. Hädrich, J. Rothhardt, M. Krebs, S. Demmler, A. Klenke, A. Tünnermann, and J. Limpert, "Single-pass high harmonic generation at high repetition rate and photon flux," J. Phys. B: At. Mol. Opt. Phys. **49**, 172002 (2016).
6. C. Benko, T. K. Allison, A. Cingöz, L. Hua, F. Labaye, D. C. Yost, and J. Ye, "Extreme ultraviolet radiation with coherence time greater than 1 s," Nat. Photon. **8**, 530–536 (2014).
7. M. Huber, W. Schweinberger, F. Stutzki, J. Limpert, I. Pupeza, and O. Pronin, "Active intensity noise suppression for a broadband mid-infrared laser source," Opt. Express **25**, 22499–22509 (2017).
8. O. Pronin, M. Seidel, F. Lücking, J. Brons, E. Fedulova, M. Trubetskov, V. Pervak, A. Apolonski, T. Udem, and F. Krausz, "High-power multi-megahertz source of waveform-stabilized few-cycle light," Nat. Commun. **6**, 6988 (2015).
9. S. Goncharov, K. Fritsch, and O. Pronin, "Amplification-free GW-level, 150 W, 14 MHz, and 8 fs thin-disk laser based on compression in multipass cells," Opt. Lett. **49**, 2717–2720 (2024).
10. J. M. Dudley, G. Genty, and S. Coen, "Supercontinuum generation in photonic crystal fiber," Rev. Mod. Phys. **78**, 1135–1184 (2006).
11. T. Sylvestre, E. Genier, A. N. Ghosh, P. Bowen, G. Genty, J. Troles, A. Mussot, A. C. Peacock, M. Klimczak, A. M. Heidt, J. C. Travers, O. Bang, and J. M. Dudley, "Recent advances in supercontinuum generation in specialty optical fibers [Invited]," J. Opt. Soc. Am. B **38**, F90 (2021).
12. M. Bradler, P. Baum, and E. Riedle, "Femtosecond continuum generation in bulk laser host materials with sub-µJ pump pulses," Appl. Phys. B **97**, 561–574 (2009).
13. A. Couairon and A. Mysyrowicz, "Femtosecond filamentation in transparent media," Physics Reports **441**, 47–189 (2007).
14. A. Dubietis, G. Tamošauskas, R. Šuminas, V. Jukna, and A. Couairon, "Ultrafast supercontinuum generation in bulk condensed media," Lithuanian Journal of Physics **57**, (2017).
15. S. L. Chin, S. Petit, F. Borne, and K. Miyazaki, "The White Light Supercontinuum Is Indeed an Ultrafast White Light Laser," Jpn. J. Appl. Phys. **38**, L126 (1999).
16. J. R. C. Andrade, N. Modsching, A. Tajalli, C. M. Dietrich, S. Kleinert, F. Placzek, B. Kreipe, S. Schilt, V. J. Wittwer, T. Südmeyer, and U. Morgner, "Carrier-Envelope Offset Frequency Stabilization of a Thin-Disk Laser Oscillator via Depletion Modulation," IEEE Photonics Journal **12**, 1–9 (2020).
17. J. J. McFerran, W. C. Swann, B. R. Washburn, and N. R. Newbury, "Suppression of pump-induced frequency noise in fiber-laser frequency combs leading to sub-radian fceo phase excursions," Appl. Phys. B **86**, 219–227 (2007).
18. S. A. Meyer, J. A. Squier, and S. A. Diddams, "Diode-pumped Yb:KYW femtosecond laser frequency comb with stabilized carrier-envelope offset frequency," Eur. Phys. J. D **48**, 19–26 (2008).
19. C.-C. Lee, C. Mohr, J. Bethge, S. Suzuki, M. E. Fermann, I. Hartl, and T. R. Schibli, "Frequency comb stabilization with bandwidth beyond the limit of gain lifetime by an intracavity graphene electro-optic modulator," Opt. Lett. **37**, 3084–3086 (2012).
20. S. Koke, C. Grebing, H. Frei, A. Anderson, A. Assion, and G. Steinmeyer, "Direct frequency comb synthesis with arbitrary offset and shot-noise-limited phase noise," Nat. Photon. **4**, 462–465 (2010).
21. T. Balčiūnas, O. D. Mücke, P. Mišeikis, G. Andriukaitis, A. Pugžlys, L. Giniūnas, R. Danielius, R. Holzwarth, and A. Baltuška, "Carrier envelope phase stabilization of a Yb:KGW laser amplifier," Opt. Lett. **36**, 3242–3244 (2011).
22. M. Seidel, J. Brons, F. Lücking, V. Pervak, A. Apolonski, T. Udem, and O. Pronin, "Carrier-envelope-phase stabilization via dual wavelength pumping," Opt. Lett. **41**, 1853–1856 (2016).
23. A. Klenner, F. Emaury, C. Schriber, A. Diebold, C. J. Saraceno, S. Schilt, U. Keller, and T. Südmeyer, "Phase-stabilization of the carrier-envelope-offset frequency of a SESAM modelocked thin disk laser," Opt. Express **21**, 24770–24780 (2013).
24. N. Modsching, C. Paradis, P. Brochard, N. Jornod, K. Gürel, C. Kränkel, S. Schilt, V. J. Wittwer, and T. Südmeyer, "Carrier-envelope offset frequency stabilization of a thin-disk laser oscillator operating in the strongly self-phase modulation broadened regime," Opt. Express **26**, 28461–28468 (2018).
25. K. Suliga, J. Sotor, and M. Kowalczyk, "Direct electro-optic phase control for carrier-envelope offset frequency stabilization in solid-state lasers," Opt. Express **33**, 21870 (2025).
26. S. Gröbmeyer, J. Brons, M. Seidel, and O. Pronin, "Carrier-Envelope-Offset Frequency Stable 100 W-Level Femtosecond Thin-Disk Oscillator," Laser & Photonics Reviews **13**, 1800256 (2019).

27. M. Hoffmann, S. Schilt, and T. Südmeyer, "CEO stabilization of a femtosecond laser using a SESAM as fast opto-optical modulator," Opt. Express **21**, 30054–30064 (2013).
28. B. Borchers, A. Anderson, and G. Steinmeyer, "On the role of shot noise in carrier-envelope phase stabilization," Laser & Photonics Reviews **8**, 303–315 (2014).
29. B. Borchers, S. Koke, A. Husakou, J. Herrmann, and G. Steinmeyer, "Carrier-envelope phase stabilization with sub-10 as residual timing jitter," Opt. Lett. **36**, 4146–4148 (2011).
30. S. Goncharov, K. Fritsch, and O. Pronin, "110 MW thin-disk oscillator," Opt. Express **31**, 25970–25977 (2023).
31. M. Kadiwala, Y. Kopp, N. Kovalenko, S. Goncharov, and O. Pronin, "High-harmonic generation directly driven with 250 MW thin-disk oscillator system," in *Conference on Lasers and Electro-Optics/Europe (CLEO/Europe 2025) and European Quantum Electronics Conference (EQEC 2025) (2025), Paper Cf_2_3* (Optica Publishing Group, 2025), p. cf_2_3.
32. Y. Kopp, S. Goncharov, G. Hehl, and O. Pronin, "Carrier-Envelope Offset Frequency Characterization of a 100 MW-Level Thin-Disk Oscillator," in *Conference on Lasers and Electro-Optics/Europe (CLEO/Europe 2023) and European Quantum Electronics Conference (EQEC 2023) (2023), Paper Cf_5_4* (Optica Publishing Group, 2023), p. cf_5_4.
33. J. Brons, V. Pervak, D. Bauer, D. Sutter, O. Pronin, and F. Krausz, "Powerful 100-fs-scale Kerr-lens mode-locked thin-disk oscillator," Opt. Lett. **41**, 3567–3570 (2016).
34. Y. Kopp, G. Hehl, K. Schwarz, and O. Pronin, "Repetition rate stabilized high power mode-locked thin-disk oscillator," presented at CF-5.5 Ultrafast Laser Technology 1 (June 24, 2025).